# Maintainability and evolvability of control software in machine and plant manufacturing - an industrial survey

Birgit Vogel-Heuser and Felix Ocker
Institute of Automation and Information Systems, Technical University of Munich, 85748 Garching, Germany
(e-mail address of corresponding author: vogel-heuser@tum.de, Fax: +49 89 289 16410)

*Abstract*—Automated Production Systems (aPS) have lifetimes of up to 30-50 years, throughout which the desired products change ever more frequently. This requires flexible, reusable control software that can be easily maintained and evolved. To evaluate selected criteria that are especially relevant for maturity in software maintainability and evolvability of aPS, the approach SWMAT4aPS+ builds on a questionnaire with 52 questions. The three main research questions cover updates of software modules and success factors for both cross-disciplinary development as well as reusable models. This paper presents the evaluation results of 68 companies from machine and plant manufacturing (MPM). Companies providing automation devices and/or engineering tools will be able to identify challenges their customers in MPM face. Validity is ensured through feedback of the participating companies and an analysis of the statistical unambiguousness of the results. From a software or systems engineering point of view, almost all criteria are fulfilled below expectations.

## 1. Motivation and Introduction

In the domain of automation, different disciplines, namely mechanics, electrics/electronics, and software, interact with and strongly depend on each other. Therefore, the success factors for cross-disciplinary development in the domain of automated Production Systems (aPS) are investigated.

The complexity of aPS, including automation hardware and automation software (latter called software henceforth), is steadily increasing. Since the proportion of system functionality realized by software is growing [1], concepts for supporting automation software engineers in handling this complexity and maintaining the developed software are required. To cope with this complexity, e.g. modular structures and variant and version management may be applied. Modularity is especially important for Industry 4.0 software, which must be easily changeable to provide additional or optimized functionality.

Manufacturing companies operating machines or plants nowadays have to meet various challenges, including small lot sizes, high variability of product types, and a changing product portfolio during the lifecycle of a machine or plant, in the following referred to as aPS [2, 3, 4]. Therefore, aPS have to support changes in their physical layout, including extensive technical updates [5], as lifecycles may last up to 50 years [6]. This paper investigates how companies from machine (MM) and plant manufacturing (PM) cope with this challenge. Existing and potential factors are identified that enable the successful exchange of software modules in aPS already in operation.

Due to high competition, times to market are also shortened for companies in the domain of aPS. This challenge needs to be tackled by reusing software modules. To do so effectively, a certain maturity of these modules is a prerequisite. A questionnaire is used to analyze which approaches for planned reuse from academia are already applied in industry and in which aspects companies still need to improve.

The application of SWMAT4aPS (software maturity for automated Production Systems) [7] resulted in various insights into the state of the art in industry concerning modularity and architecture of Programmable Logic Controller (PLC)-based software for aPS. The key contribution of SWMAT4aPS [7] was to develop a first questionnaire to measure maturity in modularity ($M_{MOD}$), testing/quality assurance ($M_{TEST}$), and start-up/operation/maintenance ($M_{OP}$) as well as the overall maturity. The results were confirmed by expert interviews and code reviews. In total, 16 companies were included that were known beforehand. However, there are some limitations to SWMAT4aPS concerning the analysis of maintainability and evolvability of aPS control software. These aspects are crucial, though, as they are prerequisites for lifelong evolution of manufacturing systems by updating their software. To gain additional insights into the state of the practice, SWMAT4aPS+ was developed, which is introduced in this paper. Comparing SWMAT4aPS with SWMAT4aPS+, all questions included in Q1 were changed due to a variety of reasons:

1. New questionnaire – all questions were modified for Q2 in SWMAT4aPS+ compared to Q1 in SWMAT4aPS.
    a) Different focus leads to other research questions: in this paper, the focus lies on maturity in software exchange during the operation phase of aPS as a prerequisite for their evolvability. This necessitated to remove/add questions and to adapt questions from Q1 related to modularity maturity.
    b) Lessons learned from SWMAT4aPS: design of multiple/single choice (MC/SC) questions, inclusion of text boxes where appropriate, and formulation of follow-up questions to get specific information.
    c) Requirements of web-questionnaires: MC/SC instead of open questions and more choices to increase precision as well as optional text boxes for a better interpretability.

2. Only questionnaire instead of questionnaire plus expert interviews: Interpretability improvement via MC/SC and consistency checks between several questions.
3. Large-scale evaluation with unknown (random, not biased) companies via web interface: additional generic questions about companies' background, evaluation of a comparison group of seven known companies.

Industry 4.0 requires flexible aPS to produce customer specific products. Since such flexibility is mostly gained through software changes, this paper provides a survey of the state of the art in the machine and plant manufacturing industry of the associated prerequisites. This paper also aspires to identify weaknesses in software engineering of aPS as a basis for further research. Within this paper, three main research questions (RQ) are addressed:

1. To allow the adaptation of a machine or plant by software: How can the exchange/update of aPS software modules be eased or enabled after acceptance test?
2. As software changes often arise from the mechanics or electrics discipline: What are success factors for cross-disciplinary development?
3. To shorten the development time and achieve a shorter time to market: What are success factors for reusable software modules?

For these research questions, several corresponding detailed questions are derived. Validity is ensured by feedback of the participating companies, a comparison with other research papers and an evaluation of the statistical unambiguousness.

The remainder of the paper is structured as follows: First, the state of the art presents related work from which research questions are derived. Subsequently, the research method SWMAT4aPS+ is described. Within the fourth section, general information about the interviewed companies as well as clusters of variables, i.e., questions, are introduced. The questionnaire's results are grouped according to the three main research questions and are presented in sections 5 through 7. Section 8 describes major influencing factors on maintainability of software modules and inferred necessary measures. The paper closes with section 9 providing a conclusion regarding the applicability of the method and an outlook on covering the identified weaknesses in future work.

**2. Related Work on Software Modularity for aPS**

The state of the art is structured into seven sections. First, a brief introduction is given to the technology of Cyber-Physical Systems (CPS) as enabler for Industry 4.0 and adaption of aPS. Next, the characteristics of aPS are delineated, including typical platforms and programming languages, to better understand the requirements and constraints of this domain. Since Model Driven Engineering (MDE) is a prerequisite for CPS, the state of the art of MDE is introduced as well as software architectures in aPS. The final three sections present an overview of selected means discussed in academia and industry to cope with aPS-specific challenges. Apart from configuration management and variant and version management, the use of hierarchy levels and standard software functions is described and the mechatronic approach is introduced, which was developed to integrate the involved disciplines. This part of the state of the art highlights selected advances academia made and introduces the RQs, which are meant to identify the gap between academia and industry.

*2.1. CPS enabling Industry 4.0*

CPS are generally defined as a "merger of cyber (electric/electronic) systems with physical things" [9]. To describe CPS, Lee et al. [10] introduced the 5C cyber-physical architecture consisting of the levels (smart) Connection, (data-to-information) Conversion, Cyber, Cognition and Configuration. CPS are expected to play a major role in design and development of future engineering systems and thus, to be a key enabler for Industry 4.0 [9]. In the production context, CPS are called Cyber-Physical Production Systems (CPPS). They especially have potential concerning autonomy, functionality, usability, reliability, and cyber security [9]. By definition, CPPS consist of interlinked aPS. According to Berardinelli et al. [11], some of the main requirements towards CPPS, and thus to interconnected aPS, are interoperability, virtualization, decentralization, real-time capability, modularity, and cross-disciplinary methods.

*2.2. Characteristics of aPS*

Complexity and variations resulting from customer-specific requirements as well as the degree of on-site changes are increasing from MM to PM [4] and in MM from standardized to special purpose machines. Therefore, three different business types are differentiated in the following, i.e. standardized machine, special purpose machine and plant manufacturers [4, 7].

The lifecycle of aPS may be divided into two main phases, at first engineering, which includes testing/quality assurance, and second, after acceptance test, operation and maintenance. The operation phase of aPS poses especially challenging requirements, as it may last up to 30 to 50 years [6]. At the same time, the start-up phase of a machine or plant shall be shortened to reduce the time to market of a new product as well as the start-up costs. As different companies from different business types face different challenges, they develop different processes and solutions. Due to the nature of plants, changes resulting from new products, unforeseen raw material or environmental conditions often have to be made on-site [4, 12]. The resulting downtime is extremely costly and should be minimized.

Such changes can often be achieved by exchanging software during a short standstill or at runtime. This paper intends to uncover the underlying prerequisites (RQ3) and success factors for reusable software modules, which are in turn prerequisites for flexibility in terms of updates that allow producing a new product (RQ1). Since complexity varies for the different business types, its influence on reusable software modules needs to be examined (RQ3.4). In case of software updates during both a short standstill or even during operation of aPS, the software quality should be as mature as possible and therefore, it is investigated how software quality is assured (RQ3.3).

Generally, a cross-disciplinary development is typical for all aPS and software changes may be initiated by mechanics, electrics or electronics. Thus, success factors for cross-disciplinary development are of great interest to the domain of aPS (RQ2).

### 2.3. Platforms, programming languages and software architectures for aPS

To understand the requirements of aPS, typical platforms, software architectures and programming languages are introduced because they constraint the applicability of usual software engineering approaches.

PLCs are the typical means in automation to control plants and machines. They are characterized by their cyclic data processing behavior and PLC cycles adhere to real-time requirements, i.e. the defined cycle times of the tasks may never be exceeded. The commonly used IEC 61131-3 programming standard for PLC consists of two textual languages – Structured Text (ST) and Instruction List (IL) – and three graphical languages – Ladder Diagram (LD), Function Block Diagram (FBD), and Sequential Function Chart (SFC).

Apart from IEC 61131-3 and PLC, also high-level/Object Oriented (OO) programming languages and more sophisticated control platforms are applied in industry. Werner [13], Bonfé et al. [14] and Vogel-Heuser [15] highlighted the benefit of OO Programming (OOP). This paper aims at identifying to what extent they are used (RQ1.3). Tool support for the OO extension of the IEC 61131-3 is now available for selected runtime environments [13]. It is assumed that usage of OO will tremendously ease reuse and modularity as highlighted by the authors mentioned above. Therefore, the relation between OO IEC 61131-3 and reusable software as well as available platforms should be analyzed because up to now only selected application examples are available (RQ1.3a, b).

As appropriate software architectures are supposed to ease maintainability and evolvability [4, 16] they will be discussed in the following. The software architecture of aPS, which contains software components and their connections, highly influences quality criteria such as changeability, maintainability or performance. An appropriate software architecture is crucial to ensure high software quality and to enable reusability.

Vogel-Heuser et al. [17] distinguished five architectural levels by analyzing the software architecture of seven companies from the MM and PM industry. These five levels of granularity are plant modules, facility modules (machines), application modules (machine parts), basic modules (e.g. individual drives/sensors) and atomic basic modules, which cannot be decomposed further. In both cases, the architectural levels can be used recursively. The higher the granularity, the higher the potential for reuse. However, high granularity also leads to higher organizational effort for combining the larger number of modules. Maga et al. [18] stated that software modules should be managed in an appropriate way according to their level of granularity.

### 2.4. MDE in aPS

Research in the field of MDE lately focused on new methods and notations to support the development of control software to ease the development, reduce cost, and support maintainability and evolvability [4, 14, 19]. However, there exists a large gap between legacy control code (i.e., PLC code) and code gained through newer MDE approaches based on the Systems Modeling Language (SysML) or the Unified Modeling Language (UML) [14] in aPS companies. To bridge this gap, both code refactoring and the building of appropriate software components are essential.

Vyatkin [20] introduced a software architecture for distributed automation systems based on the IEC 61499 standard. This approach results in software showing a composite structure and consisting of event-driven Function Blocks (FBs), which are used to describe processes. Major benefits of this approach are reduced time and effort to develop automation software, a high degree of code modularity and a high potential for reuse. These benefits are already proven through first industrial applications. However, this standard is not commonly used in industry yet and "[...] has [still] a long way in order to be seriously considered by the industry" [21]. With these approaches from academia and the huge benefit already realized for example in the automotive industry [22] in mind, the extent shall be discussed, to which MDE is used to ease adaptability of aPS (RQ1.2). The uniqueness of aPS further complicates using MDE approaches, making variant and version management all the more crucial (cp. section 2.5).

Modeling tools are not only available for software development, but also for the other disciplines involved in the development of aPS. Mechanical engineers express a lot of their expertise in CAD data, while electrical engineers use tools ranging from the generic Excel to domain specific ECAE tools, like EPLAN EE respectively P8 or Zuken E³. Combining the models of the different disciplines involved allows for extensive simulations. These in turn can be used to validate machine or even plant designs, which reduces the effort for the subsequent testing. This thought was already developed in the 90s (cp. e.g. [23]) and a variety of approaches have been proposed since [24, 25, 26].

### 2.5. Configuration management, variant and version management

As introduced above, a variety of renowned researchers including Egyed, Prähofer, Vogel-Heuser, Fay, Tichy and Schaefer (cp. e.g. [4, 27, 28]) identified variant and version management as a prerequisite for a proper software engineering in the aPS domain and, therefore, also the enabler of software updates in the operation phase. In aPS, new variants (VB) are often derived from existing variants (VA). Due to parallel operation with different machines for different customers at different sites, faults in variant VB may occur before they arise in the initial variant VA. Additionally, the plant or machine operator may have adapted the original software. Nonetheless, faults must be fixed via updates (RQ1.1a) in both variants, necessitating knowledge about variants and versions as well as configuration management. This paper investigates whether companies have sufficient knowledge about variants and versions for updating (RQ1.1) as well as the influence of variant and version management tools on management of SW (RQ3.1). Two ways are explicitly investigated for gaining knowledge concerning variants and versions as they are in use on site. These are code sharing techniques (RQ1.1b) and the acquisition of the customer's SW status by technicians (RQ1.1c).

Stallinger et al. [29] developed a process reference model for reuse in industrial engineering, which they validated for the software engineering domain. They distinguish four reuse maturity stages, namely chaotic, systematic, domain-oriented and strategic.

Vogel-Heuser at al. [4] provide a survey on variability and feature modeling in the aPS domain based on the work on software product lines of Pohl et al. [30] and feature models (e.g. Benavides et al. [31]). They found that fundamental methods supporting variability of modular aPS are still limited. To cope with the interdisciplinary nature of aPS, these methods should be adapted to meet the needs of all disciplines involved and should be linked to relevant domain-specific models requiring appropriate tool support. While clone detection and code management are common in software engineering, so far only simple versioning is available for aPS development platforms and IEC languages. Clones in aPS are not only software clones but also mechanical and electrical/electronic clones embedded in different engineering tools [4]. Such engineering tools range from UML and Matlab/Simulink models to tools from electrical engineering even to rudimentary Excel. To cope with variability, product lines and feature models are important issues [4, 32], but cross-disciplinary approaches are still in their beginning.

In industry, many integrated platforms still exist that are based on a cloning approach for creating new product variants. Antkiewicz et al. [33] suggest to address the challenge of migrating such an integrated platform into a central platform with the virtual platform strategy. It covers seven governance levels, ranging from ad-hoc clone and own (level L0) to product line engineering (PLE) with a fully-integrated platform (level L6). As the results of the first questionnaire [7] showed, the method copy, paste and modify is still widespread. In addition, the governance level was below L4 for all four case studies that were analyzed in detail by Vogel-Heuser et al. [7]. Considering the broad range of levels and these first insights, it is further investigated to which degree code configuration from engineering tools is realized in industry (RQ3.6). An alternative to the method copy, paste and modify is the reuse of program organization units (POUs) collected in libraries. The release process of these POUs in industry is investigated (RQ3.2).

A prior survey of the author in 2014 [34] in the aPS domain focused on the implementation of variant and version management (VVM). It revealed that 30% of the companies implemented VVM and 30% implemented VVM partially within a single discipline but across disciplines, only 10% implemented VVM and 20% partially implemented VVM. Arvanitou et al. [35] examined the state-of-research of design-time quality attributes in general software engineering by conducting a mapping study including 154 papers revealing that maintainability is the most frequently examined high-level quality attribute, regardless of the application domain or the development phase. They found metrics like Lines of Code (LOC) and Cyclomatic Complexity to be relevant. Weyrich et al. [36, p. 189] name Halstead complexity measures as other possible metrics. Capitán et al. [37] used such metrics for aPS code to measure code size, complexity, decomposability, communication, module size and hierarchy. However, they did not deliver meaningful results for code evolution. Additionally, companies participating in Q1 were unable to provide LOC or similar measures. Concluding, there is still no reliable and accessible measure for complexity in aPS.

### 2.6. Hierarchy levels and standard functions for diagnosis and fault handling

The well-known standard ISA-88 [38] provides an appropriate hierarchy of modules [39]. This hierarchy can be mapped to the granularity levels presented by Vogel-Heuser et al. [17]. Thereby, a basic module corresponds to a control module while a plant module is equivalent to a process cell. Depending on the hierarchy level, modules implement typical standard functions. This supports software architectures and eases changes and adaptation. A standard widespread in the food and beverages industry is PackML. It includes OMAC state machines [40], which are used to realize operational consistency within a packaging line. In automation, examples for such standard functions are diagnosis, i.e. fault detection and fault handling [12, 41]. The questionnaire investigates to which degree standard functions and hierarchy levels are realized in industry (RQ3.5).

### 2.7. Mechatronic approach

The collaboration of different disciplines, mostly mechanics, electrics/electronics and software, still poses a major challenge in the development of mechatronic products [42]. This is due to a variety of interdependencies that are not considered by domain-specific tools and thus necessitate exchange. Hence, the tools shall be identified that are common for information-exchange among the involved disciplines (RQ2.2).

Langer et al. [43] investigated technical changes in 94 German industrial companies from a production development point of view. They identified initiators of change and communication channels among the involved stakeholders. Most important is direct communication, taking place in meetings or phone calls. This can be either formal or informal. In case of greater distance, email or document based messages are used more often. They identified a surprisingly low usage of software solutions like Product-Lifecycle-Management (PLM) software. This underlines the importance of formal or informal meetings to discuss changes during the design process. Research and Development (R&D) in product development may be rated similarly to the design department in a plant manufacturing or special purpose machine manufacturing company.

There are various development processes for complex products that can be applied to the development of mechatronic products, including machines and plants. These range from established ones such as the V-model to more recent ones like Scrum. Scrum helps cross-functional teams to accelerate development by breaking down a complex project into smaller chunks, which are realized in so-called sprints in case of deviations from the project schedule. Instead of adhering to regular intervals, meetings are held according to necessity, usually daily, during these sprints [44].

In addition to these generic development processes, more specific approaches were developed for coupling especially the disciplines mechanics, electrics/electronics and software. Mechatronic System Models that integrate discipline-specific models have been proposed to cope with this challenge [42]. In this work, especially the validation of design concepts in early phases is addressed. Another way for designing mechatronic systems is presented by Zheng et al. [45]. Here, an interface model is used for integrating an extended V-model and a hierarchical design model. Similarly, Mensing et al. [46] reuse existing discipline-specific models and interface them to different abstract models, i.e. an adaptation of the SysML requirement diagram, of the system. Shah et al. [47] use a multi-view modeling approach, so that established tools may be maintained. By use of a common system model and transformations to and from the discipline-specific models, the different views are interwoven, resulting in a model integration framework. Thramboulidis [1] proposed the 3+1 SysML view-model to integrate the parts of the three domains prevalent in automation. In accordance with the model-integrated mechatronics paradigm [48], mechatronic components are created, which use mechatronic ports as interaction points and may be composed to an aPS. However, not all involved disciplines, namely mechanics, electrics and software, are relevant for all components [49]. This means, there may exist components that are e.g. purely software and have no mechanical counterpart, which makes a universal 1:1:1 mapping impossible. Within this paper, it is investigated whether cross-disciplinary modules are used, and if so, which ones (RQ2.3). A SysML-based modeling methodology for supporting the design process of complex manufacturing systems was developed by Bassi et al. [50]. Hereby, a hierarchy of models is used to represent three levels of abstraction. The highest level serves the understanding of functions and interdependencies, while the specific executable models on the lowest level may e.g. be used for simulations. Kernschmidt et al. [49] introduced the approach SysML4*Mechatronics* [51]. It adapts the modeling language SysML by use of a port-concept to provide the means for inter-disciplinary modeling. In this way, the change influences due to dependencies among the different disciplines involved may be assessed. Hereby, the focus is on changes caused by new requirements or forced innovations. The latter are induced by the different lengths of lifecycles in the disciplines [49]. By use of a SysML-based design pattern, the information from different levels of abstraction and different disciplines can be connected and a functional modularization may be conducted [52]. In this way, change influences can be evaluated, fulfilment of requirements may be assessed, and reusability is enhanced, which allows for a drastic shortening of development cycles. As argued by Feldmann et al. [53], an increased demand for adaptability and flexibility in automation leads to an increase in complexity. As the manual selection of appropriate components from large component libraries is inefficient, a vision for an automatic synthesis of manufacturing system designs is presented. Emphasis is especially put on the compatibility of components, since this is crucial for a functioning system. With this variety of approaches from academia in mind, it is investigated how cross-disciplinary development is realized in industry (RQ2.1).

## 3. Research Method

The research goal addressed in this paper is to gain deeper insights in the state of the art in software engineering of aPS and to identify weaknesses as a basis for further research. Three research questions were identified that shall be addressed. To support adaptability along the aPS life cycle, new functions need to be implemented in aPS that are already in use. Therefore, RQ1 addresses prerequisites to enable or support the exchange of software modules for operating aPS (cp. section 5). Since aPS are mechatronic systems, software is strongly related to mechanics and electrics. To achieve manageable software exchange as discussed in RQ1 the entire cross-disciplinary development needs to be set up accordingly, which is addressed by RQ2: What are the success factors for cross-disciplinary development in general? (cp. section 6) Additionally, in a microscopic view into the software discipline, RQ3 addresses the success factors for software modules (cp. section 7). These three research questions and the associated detailed research questions are listed in Table I.

**TABLE I**
**Research Questions**

| Research Questions | Detailed Research Questions |
| --- | --- |
| How can the exchange of aPS software modules be enabled at runtime? (RQ1) | Do companies have sufficient knowledge about variants and versions of SW for updating? (RQ1.1) |
| | Frequency of software updates (RQ1.1a) |
| | Code sharing strategies (RQ1.1b) |
| | Acquiring software status (RQ1.1c) |
| | To what degree do companies use the potential of modeling tools to ease adaptability? (RQ1.2) |
| | Which high-level programming languages and more sophisticated control platforms are used? (RQ1.3) |
| | Usage of OO (RQ1.3a) |
| | Dependency between OO and IEC (RQ1.3b) |
| What are success factors for cross-disciplinary development? (RQ2) | How is cross-disciplinary development realized? (RQ2.1) |
| | Which tools are used for supporting information exchange among different disciplines? (RQ2.2) |
| | Usage of variant mgmt. tools (RQ2.2a) |
| | Disciplines included in version mgmt. (RQ2.2b) |
| | Are cross-disciplinary modules used? (RQ2.3) |
| What are success factors for reusable software modules? (RQ3) | How big is the variant and version mgmt. tools' influence? (RQ3.1) |
| | How are library blocks released? (RQ3.2) |
| | How is quality assured? (RQ3.3) |
| | What is the influence of complexity? (RQ3.4) |
| | Which standard functions are most used? (RQ3.5) |
| | Is code configuration from engineering tools realized in industry? (RQ3.6) |

mgmt. = management.

The benchmark process SWMAT4aPS+ (cp. Fig. 1) takes the preparation step from SWMAT4aPS [7] and adapts the questionnaire firstly by adding questions to analyze selected aspects in more detail and secondly rephrases questions for a better understanding and the limited question types available at the used web based platform. The latter is even more important for a web-based questionnaire without the opportunity for participants to ask if questions remain unclear or ambiguous. In the experimentation part, the pre-processing had to be adapted due to other identified phenomena and the reporting was enlarged accordingly to highlight relations between questions' ratings. The approach realized in this survey does not include an expert analysis of code as in [7], though.

### 3.1. Questionnaire

The SWMAT4aPS+ questionnaire is limited to 52 questions, including sub- and follow-up questions, due to the restricted time of employees to spend on extra work (max. 15-30 minutes). The questionnaire is divided into five sections: questions on company and equipment (15 questions), approaches in development (7 questions), software and reusability (19 questions), quality assurance (5 questions) and finally handover to the customer and commissioning of the plant (6 questions). In the following, specific questions are indicated by (#number of question). Different types of questions were used, i.e. select from options (single or multiple choice) ranging from 6 to a maximum of 11, percentages (#3.17) and free text (#2.4, #3.5). The full questionnaire is available online [54].

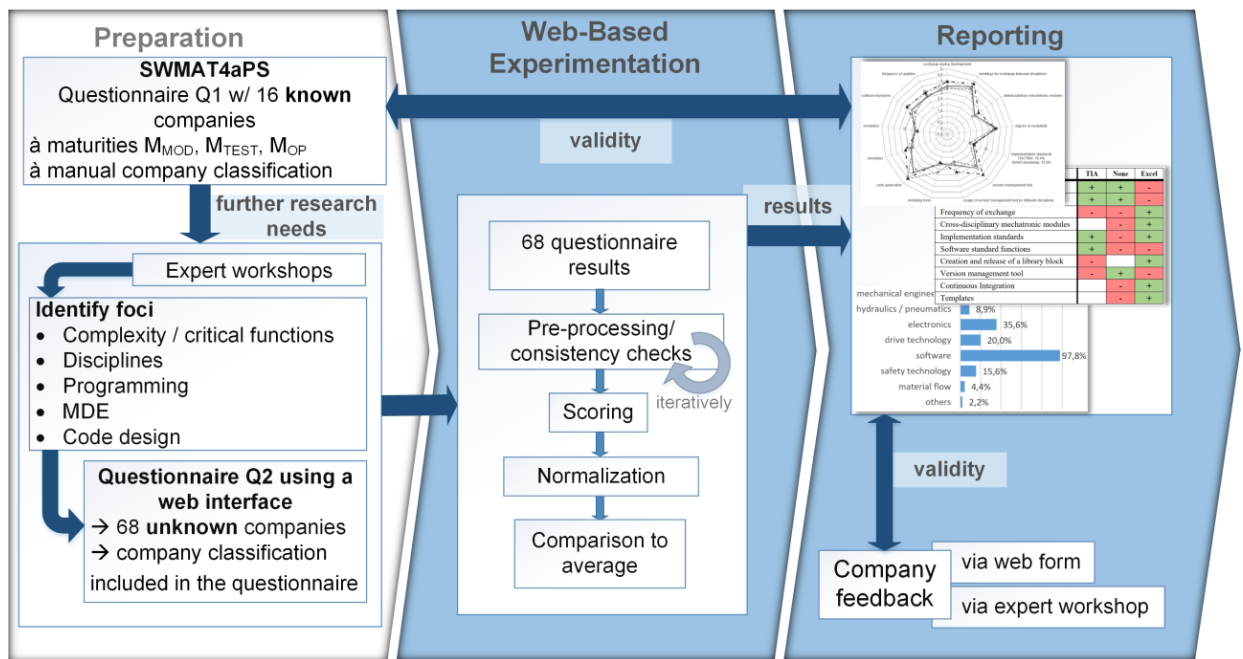

Fig. 1. SWMAT4aPS+ simplified benchmark process.

In the following, the changes of Q2 compared to Q1 are discussed in detail, as listed in the introduction. Comparing Q1 and Q2, it becomes apparent, that all questions were modified in some way. As stated in 1.a), different foci lead to other research questions. Q2 is directed at software exchange during operation and its prerequisites as a basis for evolvable aPS. Therefore, half of the questions from Q1 was removed because they are out of scope, e.g. Q1 #5.1.2 "In which phase are code reviews used?" On the other hand, new ones were added (approx. 30%) to address the RQs specified for this paper, e.g. Q2 #3.2 "Do you use the object-oriented amplification of the IEC 61131-3 programming languages?" Given the assumption that modularity in design is key to enable software exchange during operation, some questions from Q1 that address $M_{MOD}$ and $M_{OP}$ were used and elaborated on further. To adapt questions to the new RQs, about 25% of questions from Q1 were changed significantly in meaning for Q2. An example is Q1 #3.1.1 "Which programming languages are used in your company?", which was changed to "Which programming languages are used for control software in your company?" as Q2 #3.1. Lessons learned from SWMAT4aPS were also realized (1.b). The results, especially those of free text questions, from Q1 proved to be a valuable basis for designing multiple and single choice questions. Free text questions in Q2 were included to find new indicators for further analysis. Additionally, follow-up questions were included in Q2 to get information that depends on certain circumstances. Q2 was also adapted to the requirements of web-questionnaires (1.c). This web-format of the questionnaire makes a detailed discussion of the results with the companies impossible, as it was conducted in prior industry surveys [7, 8]. Hence, questions need to be more precise to avoid ambiguities and SC and MC questions are preferred over open ones. Among the quarter of questions that were kept from Q1 for Q2, open questions were replaced by SC/MC questions, additional possible answers were included, or optional textboxes were added to improve interpretability.

Another major difference between Q1 and Q2 is that only a questionnaire with a comparison group was used instead of a questionnaire plus expert interviews (2). To still get reliable results, open questions were replaced through precise MC/SC questions, and text boxes were included for additional information. Additionally, consistency checks were conducted between several questions.

Finally, SWMAT4aPS+ is a large scale evaluation with unknown (random, not biased) companies via web interface (3). In total, 68 companies completed the web-based questionnaire that was spread by newsletters from a publishing house. Apart from seven companies, which served as a comparison group, the authors did not have prior knowledge of the participating companies. Hence, the questions had to be tailored to the requirements of a large-scale evaluation and the fact that the authors were not able to enquire further if questions were answered inconsistently or ambiguously. Also, free text questions had to be reduced and questions had to be included to classify companies according to their market segment and type of business.

### 3.2. Experimentation

The individually answered questionnaires were classified into the three categories MM, PM and others, which are mostly engineering companies.

*Companies included in the questionnaire*: In order to capture a broad database and to ensure anonymity, the questionnaire was provided online via the market research group of the publisher *Verlag Heinrich Vogel*. A German speaking community was addressed via newsletters and web pages, which is interested in Embedded Systems and software engineering in MM and PM. Additionally, a comparison group was formed by sending individual links of the questionnaire to selected companies. This allowed identifying these specific companies and their answers. In contrast to SWMAT4aPS, knowledge about most companies and persons answering the questionnaire is still limited, though. Based on the typical readership group of the publisher, two assumptions can be made. Regarding their role in the company, two main groups of readers were identified: 63% of readers belong to R&D while 10% are technical management. Regarding the position, 15% belong to the CTO or management level, 14% are department heads, 17% are group or project managers and 50% are on the engineering level. Additionally, dedicated links were provided to seven companies that cooperate closely with the institute. For these companies, feedback meetings with presentations and discussions of the results were realized.

*Scoring*: The scoring was developed iteratively based on Q1 using a six value grading schema (from 1 (worst) to 6 (best), or from 0 to 5). When the results for the different questionnaires were available, this scoring was slightly adapted with regard to free text answers, e.g., in case additional tools were mentioned for continuous integration (#3.13) or variant management (#3.14). That way, these tools were also considered besides the provided MC/SC options. The exemplary scoring for the number of disciplines in modularization reveals the challenge in comparing different department structures. Possible answers are mechanical engineering/mechanics, hydraulics/pneumatics, electrical engineering/electronics, drive engineering, software, safety engineering, material flow, and others. If the three key disciplines mechanical, electrical and software engineering are included, the highest score was assigned (5). The highest score was also assigned in case more than three disciplines out of the selectable choices are involved (cp. Table II).

TABLE II
**Scores for the Question "Which Disciplines are Modularized in Your Company?" (#2.7)**

| Evaluation criterion | Score |
| --- | --- |
| mechanical engineering / mechanics + electronics / electrical engineering + software or more than 3 answers | 5 |
| exactly 3 answers | 4 |
| 2 answers | 3 |
| 1 answer | 1 |
| no answer | 0 |

Based on these individual scores, also the overall maturity level of each company can be calculated (cp. [7]). This is achieved by simply dividing the sum of all scores by the sum of all achievable scores.

*Consistency checks*: The answers were checked manually for potential inconsistencies. This is enabled through similar questions, for example #3.8 (percentage of library elements in average software project) and #3.17 (composition of average software project also including libraries).

### 3.3. Reporting and feedback

The graphical representation from SWMAT4aPS has been developed further and was presented to the seven known companies and groups of companies during different workshops. The feedback was throughout positive, confirming the insights gained from the questionnaire. The results have been rated as correct, beneficial and applicable to identify strengths and weaknesses. Further suggestions to measure complexity were made and will be included in the next questionnaire. A challenge during the questionnaire's design and evaluation was to quantify the answers and to normalize the ranges for enabling comparisons. This was achieved through the scoring as described above, which was approved by experts both from academia and from industry during workshops. The methodology to design and approve the questionnaire is thus the same as the one used for SWMAT4aPS. The basic idea is to define several levels of maturity/quality for the various criteria (in each question) by experts and to assign the values 0-5. To the best of our knowledge, this is the usual proceeding in prioritization in systems engineering if different design alternatives need to be compared regarding their quality [55]. In addition, the abstraction due to the standardized scoring had to be taken into account during the evaluation to ensure no information was lost. Even though the option of free text enabled additional insights, its evaluation was especially laborious and required experience in the domain of automation.

### 3.4. Threats to validity

To avoid a systematic bias, the authors followed the guidelines of Runeson et al. [56] to construct validity and reliability of the case study as described in more detail in prior work [7]. In the following, some aspects for this survey are summarized. Validity of the approach, i.e. of the companies' self-assessment, is ensured in several ways. To avoid a systematic bias specifically in the questionnaire's design, it was discussed with experts from both academia and industry in workshops. In addition, the state of the art in academia was considered, as presented in the related work section, and the authors were able to build on lessons they learned from the first questionnaire within SWMAT4aPS. The first questionnaire and its feedback helped tremendously with formulating precise questions and providing appropriate answering choices.

Concerning the participants, a potential bias was reduced by publicly providing the questionnaire through a web-interface via an independent publisher, the *Verlag Heinrich Vogel*. To avoid errors in the evaluation, the companies' answers were checked for inconsistencies. An analysis of the comparison group's maturity in relation to the average maturity further confirmed the assessment. Additionally, the results were compared with the insights gained from previous research (comparison with results gained through SWMAT4aPS/Q1, which assessed 16 companies, cp. Fig. 19) and other studies where applicable [7, 8, 31, 57]. Even though these measures were taken, the authors acknowledge that the surveyed companies may still be a convenience sample. This is, because only data is available from companies that were willing to complete the survey. This threat is accepted, though, as it cannot be overcome with reasonable effort.

Finally, the validity of the interpretation (cp. column "evaluation results" in Table III, IV, and VIII) is assessed based on the standard deviation and the number of answers (cp. column "validity of results" in Table III, IV, and VIII). This means, that the expressivity of the data was assessed in terms of unambiguousness. Concerning the standard deviation, two corridors are introduced around the standard deviation resulting from an even distribution. The narrow corridor covers all values that deviate less than 4% from this value (rating of 3), while the wider one covers all values deviating less than 8% (rating of 2). A cluster/two clusters are considered relevant if the standard deviation differs more than 8% from the one of an even distribution (rating of 1). Results considered less obvious due to the absence of clusters are marked with an "SD" in Table III, IV, and VIII. The number of answers is considered insufficient (rating of 3) if less than 30 answers are available and questionable (rating of 2) if less than 60 answers are available. Questions to which this applies are marked with an "A" in Table III, IV, and VIII. Otherwise, the number of answers is sufficient (rating of 1). The overall rating of a result's validity is based on the sum of these two ratings for the standard deviation and the number of answers. If this sum is below or equal to two, a green rating is assigned in Table III, IV, and VIII, signaling valid results. A sum of three is considered as not reliable and is thus marked orange. Results with sums greater than three are considered as the most unreliable and are marked red.

## 4. General Information about Interviewed Companies and Clusters of Variables

Based on the general information of the first part of the questionnaire, the participating companies were classified into different groups. The first dimension of classification is the industrial sector (#1.1, cp. Fig. 2). 73.5% of the companies operate in more than two industrial sectors (8.8% chose two, 7.3% chose three and some even five). Second, the companies were classified according to their type of business: 72% of the companies assigned themselves only to one type of business (special purpose, standardized machine, PM or others), but 23.5% chose two types (#1.3). Consequently, companies operating in more than one industrial sector or more than one type of business are included in all respective categories. Therefore, standardized and special purpose machinery are summarized in one category.

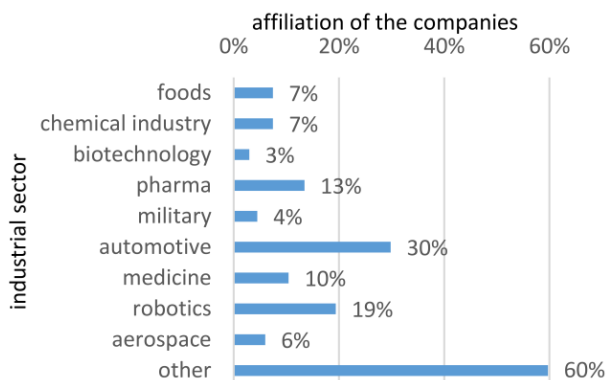

Fig. 2. Industrial sector of participating companies. (#1.1)

Similar to [37], the companies are grouped according to their number of employees as an indicator for their size (cp. Fig. 3).

To describe complexity of software, the average size of the realized software projects was measured via the number of POUs. In general, machines have less POUs than plants, whereas the number of POUs varies greatly in both cases (cp. Fig. 4). Thus, a company's type of business, i.e. MM or PM, correlates to this indicator.

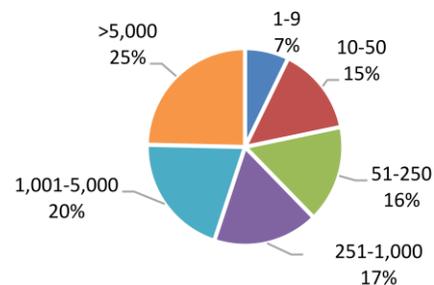

Fig. 3. Number of companies' employees. (#1.5)

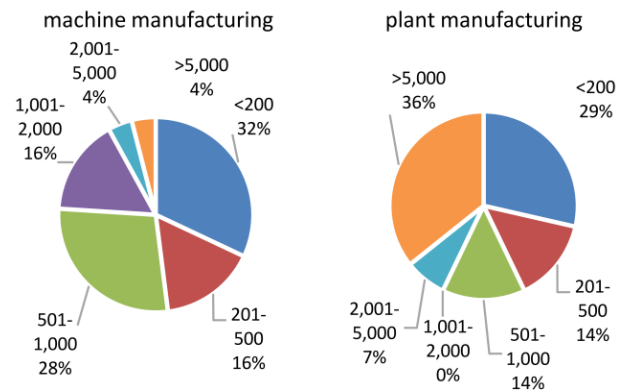

Fig. 4. Size of software projects – average number of POUs for machine and plant manufacturing, respectively. (#1.11)

At first, a factor analysis has been conducted to check if there exist variables with a similar response behavior. Such principle components would group the different variables of the questionnaire to an overall latent structure. In this way, we searched for main factors influencing maintainability and evolvability of control software. This analysis revealed six approaches listed in the following (description, particular variables):

(1) *classical software engineering approach*: frequency of meetings (#2.2), specification of requirements (#2.3), standard functions (#3.7), tested scenarios and guidelines for test creation (#4.2)
(2) *advanced software engineering approach*: object-oriented amplification of the IEC 61131-3 (#3.2), continuous integration (#3.13), automated configuration of the control software (#3.15)
(3) *modularity in engineering*: modularized disciplines (#2.7), implementation of interfaces (#3.6)
(4) *version management*: version management tool (#3.10), change tracking work with versions (#3.12)
(5) *variant management*: disciplines exchanging views during development (#2.1), variant management tool (#3.14)
(6) *modeling and library blocks*: modeling tools (#2.5), percentage of library blocks (#3.8), release of library blocks (#3.9), quality assurance (#4.1), simulation (#4.3)

However, the Kaiser-Meyer-Olkin (KMO) test measurement [58], which has been performed on this data set in parallel, reveals a low separation effect of the components and therefore a low accuracy and interpretability. In consequence, no latent overall structure of the influence on maintainability and evolvability of control software could be identified unambiguously. In other words, companies do not strictly adhere to one of those six approaches, but realize a mixture instead.

Therefore, the effect of each variable on the different research questions is investigated independently, which leads to more interpretable results.

## 5. Prerequisites for Updating Software Modules at Runtime (RQ1)

Downtime is costly in manufacturing. Therefrom arises the desire to fix bugs, to implement additional or optimized functionalities or to learn from other plants. Additionally, regular software-updates may be necessary if new products or processes are to be produced on existing plants. Software-updates in already operating plants require exact knowledge about the implemented application software's variant and its version as well as its interfaces.

### 5.1. Knowledge about implemented software variants and versions (RQ1.1)

Companies running a plant often force the PM to provide the source code. This enables them to make modifications to avoid downtime or to optimize the plant on their own after an acceptance test. In this case the PM does not necessarily know how the software has been changed or why and therefore which software actually runs on the plant. This aggravates the software updates because the changes need to be evaluated and if they are reasonable incorporated in the update. As discovered by Vogel-Heuser et al. [7, H3.3], the average maturity concerning tracking of changes is below 60% for MM companies.

To get a better understanding of software updates in aPS, the frequency of regular updates is analyzed first (a). Next, different code sharing strategies are compared (b) and finally the strategies to acquire a new software status from the supplier are discussed (c).

*Frequency of software updates (RQ1.1a)*

Up to now, MM and PM conduct remote software updates only at long intervals (cp. Fig. 5). However, the majority of companies does realize remote software updates, resulting in an orange rating (cp. Table III).

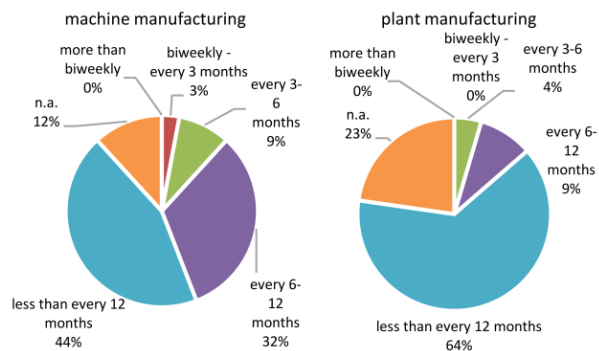

Fig. 5. Frequency of software updates. (#5.5)

*Code sharing strategies (RQ1.1b)*

In both MM and PM, 18% of the customers receive the complete software source code. However, the majority of MM companies does not supply any software code (58%) compared to 41% of the PM companies. Parts of the software code are provided by 24% in MM and 36% in PM (#5.3). Assuming that customers adapt their software if they get the source code, the status of the software is often unknown to the supplier, at least not without retrieving the newest version from the customer. This means that code-sharing strategies are not fully employed, even though a few companies manage to realize them. This leads to an overall orange rating (cp. Table III).

*Acquiring software status (RQ1.1c)*

The service department (#5.6) often gains knowledge about the current software status after the acceptance test. In MM 22% (PM 24%) of the companies get the latest software status during acceptance tests. In case a service technician is on site, 38% MM (PM 31%) of the companies get the recent status. Only 13% MM (PM 24%) receive this information on a regular basis via remote access. Taking the importance of knowledge about the current software status into account, this is a large backlog (red rating, cp. Table III).

Summarizing the aspects (RQ1.1a-c), companies which have to deliver their source code partially or entirely potentially lack the knowledge (more than 50% PM resp. 40% MM), which software is running on their plants because aggravating software updates are rare and performed regularly by less than 25%. For a significant high percentage of PM and MM the status of the software is insufficiently clear (cp. Table III orange-red dot) and therefore updates, due to customer wishes or contracts, as required for I4.0 are too risky.

### 5.2. Easing adaptability by using modeling tools (RQ1.2)

MDE is discussed since long in academia and is assumed to be increasingly implemented in aPS. The actual industrial status of MDE and its tools was analyzed (cp. Fig. 6, #2.5), revealing extensive deficits. There are differences in between MM and PM as well as among different industry sectors, for example, EA is used by 15.8% in MM and by 42.9% in PM. Existing models also enable simulation in early phases and, therefore, testing and quality checks, too. Overall, the usage of MDE approaches is still below expectations, since 44.1% in MM and 33.3% in PM do not use any.

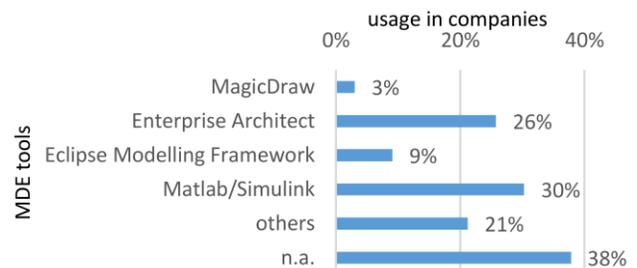

Fig. 6. Modeling tools. (#2.5)

As continuous integration (CI) is an MDE based approach to improve reactivity to changes, the dissemination of CI in aPS is of interest (#3.13), too. 52% of the participating companies do not use CI, 34% use it partially and 14% by default. No differences regarding usage of CI across industrial sectors could be identified, but regarding disciplines, nearly 100% of the companies use CI in software engineering. This leads to the conclusion that cross-disciplinary CI is still in the beginning (cp. Table III red dot).

In theory, simulation has been used for long to test software before the machine or aPS hardware is available (cp. Sec. 2.4). The degree, to which this method has reached industry is de-picted in Fig. 7. As can be seen, only 16% of all companies use simulation testing extensively, which shows that it is still far from being the standard process.

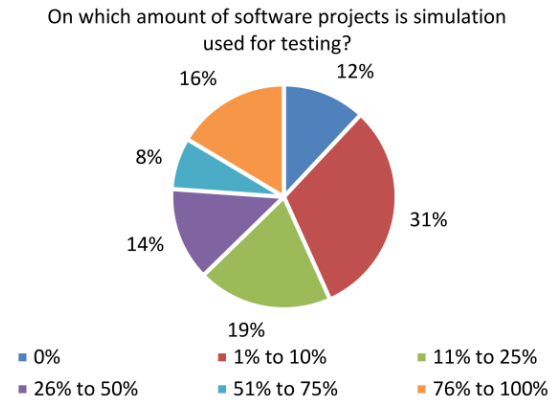

Fig. 7. Usage of simulation for testing. (#4.3)

### 5.3. Usage of high-level programming languages and more sophisticated control platforms (RQ1.3)

Besides the five classical IEC 61131-3 programming languages also high-level languages and modeling and simulation languages like Matlab/Simulink are in use. It is assumed that high-level programming languages as well as more sophisticated control platforms will ease modularity as a prequisite for adaptability. The high percentage of high-level programming languages (cp. Fig. 8) can be explained by programming of Human Machine Interfaces (HMI) and Manufacturing Execution Systems (MES), included unintentionally in the answers, which was realized during the feedback discussion with industrial partners. Therefore, in future work, this question needs to distinguish between control software and HMI, MES or other IT related applications.

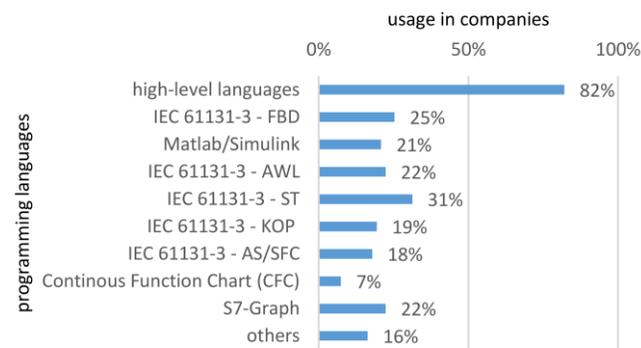

Fig. 8. Percentage of programming languages in use. (#3.1)

RQ1.3a addresses the usage of OO as one potentially promising paradigm to structure software and ease reuse. In general, not differentiating the hardware platform used for control, 42% of the companies do not use OO IEC 61131-3 at all, 48% partially and 10% by default. This means there is still a large potential for improvement for most companies assuming that OO is at least one possible way to improve reuse (orange point, cp. Table III).

Analyzing the dependency between OO IEC and the used control platform (RQ1.3b) shows, that the distributions for Embedded Systems and PLCs are almost the same, with approximately 10% using OO IEC by default, 40% using it partially and 50% not using it at all. For PC-based systems, OO IEC is applied more often partially, though (65%), while the percentage of OO IEC default use is the same with about 10%. This leaves a relatively small share of 30% of PC-based platforms that do not use OO IEC at all.

Additionally, dependencies between the used controller brand (cp. Fig. 9), the company's industrial sector and the engineering process/programming paradigm was assumed to exist as indicated by some companies and markets. Up to now, no such relations could be revealed because most MM and PM companies use more than one controller brand. This dependency needs to be analyzed in future research.

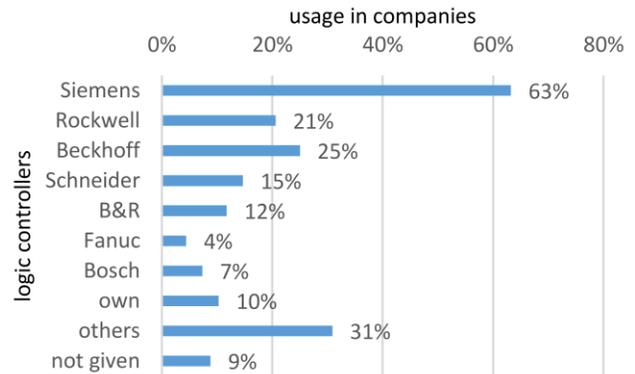

Fig. 9. Types of logic controllers in use by brand. (#1.8)

Summarizing RQ1.3a-c, one can deduce that potentials for improvement, i.a. OO, exist for MM and PM. This is expressed in an orange rating for RQ1.3 (cp. Table III).

## 6. Success Factors for Cross-Disciplinary Development (RQ2)

Within this section, the factors that are mainly responsible for successful cross-disciplinary development are investigated in detail. These success factors are information exchange and cooperation among different disciplines, supporting tools and use of cross-disciplinary modules. Variant and version management are distinguished concerning the tools.

### 6.1. Regular meetings to support information exchange in between different disciplines (RQ2.1)

Different success factors are already known for cross-disciplinary development, for example the communication between different disciplines. Such communication may be realized via meetings, and/or with technical support tools or emails and written documents ([43], cp. also Sec. 2.2). In the following, results to these strategies will be discussed for both the participating disciplines in and the frequency of meetings.

Mechanics, electrics/electronics and software are the mandatory disciplines to be involved in the development of aPS. Nevertheless, depending on the type of machinery, additionally, also other disciplines like hydraulics and pneumatics need to be involved and may be organized in different groups or even departments in different companies. Management and / or project leaders may be involved, too. The distribution of disciplines of the participating persons in the questionnaire and how these different disciplines are represented in meetings is shown in Fig. 10.

TABLE III
Evaluation of Research Question 1

| Research Questions | Detailed Research Questions | Findings | Evaluation results | Validity of results | Relevant Section |
|---|---|---|---|---|---|
| How can the exchange/update of aPS software modules be eased or enabled after acceptance test? (RQ1) | Do companies have sufficient knowledge about variants and versions of SW for updating? (RQ1.1) | high percentage of code sharing, long intervals in acceding software status and rare remote access | 🔴🟡 | 🟡 | 5.1 |
| | Frequency of software updates (RQ1.1a) | Long intervals: 44% MM, 64% PM less than every 12 months | 🟡 | 🟢 | 5.1 1) |
| | Code sharing strategies (RQ1.1b) | 58% of MM and 41% of PM do not supply any source code to customers | 🟡 | 🔴 SD | 5.1 2) |
| | Acquiring software status (RQ1.1c) | Only 13% MM and 24% PM use remote access | 🔴 | 🔴 SD | 5.1 3) |
| | To what degree do companies use the potential of modeling tools to ease adaptability? (RQ1.2) | 44% MM; 33% PM do not use modeling tools at all 52% don't use CI at all | 🔴 | 🟢 | 5.2 |
| | Which programming languages and more sophisticated control platforms are used? (RQ1.3) | 82% of companies use high-level programming languages | 🟡 | 🟢 | 5.3 |
| | Usage of OO (RQ1.3a) | 42% don't use OO IEC at all | 🟡 | 🟢 | 5.3 |
| | Dependency between OO IEC and used platform (RQ1.3b) | Partial use of OO is most widespread on PC-based platforms | 🟡 | 🟢 | 5.3 |

🔴 = low maturity / validity, 🟡 = medium maturity / validity, 🟢 = high maturity / validity,
SD = answers considered less obvious due to the absence of clusters, A = number of answers considered insufficient (< 30) / questionable (<60)

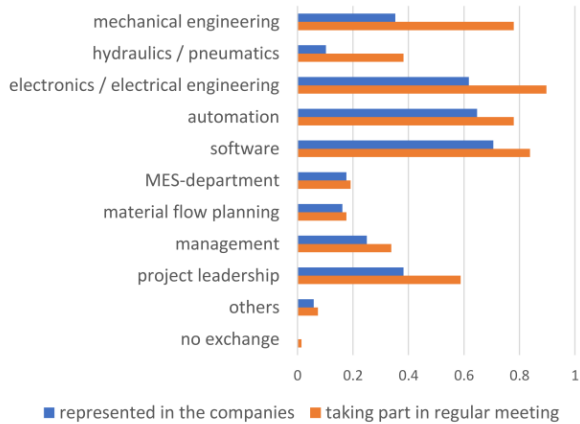

Fig. 10. Disciplines of the participating people in the questionnaire (#1.2) and disciplines attending regular team meetings (#2.1)

Fig. 10 underlines that all disciplines represented in the questionnaire's results are taking part in the meetings (green rating in Table IV). Highest participation is given from electronics/electrical engineering followed by software and automation on the same level as mechanical engineering. One could assume that electronics/electrical engineering is the connecting discipline for collaboration, but this needs to be proven in future work. The frequency of the meetings corresponds to the duration of the project. Projects in PM often last longer than in MM and, therefore, the meeting frequency decreases from weekly to biweekly. The percentage of only sporadic meetings is nearly the same for both and with 25% quite high (cp. Fig. 11). This may result from methods like SCRUM, which adjust the meeting frequency according to the necessity of sprints. Because all disciplines are attending in meetings and the frequency correlates to the project duration RQ2.1 is rated green in Table IV.

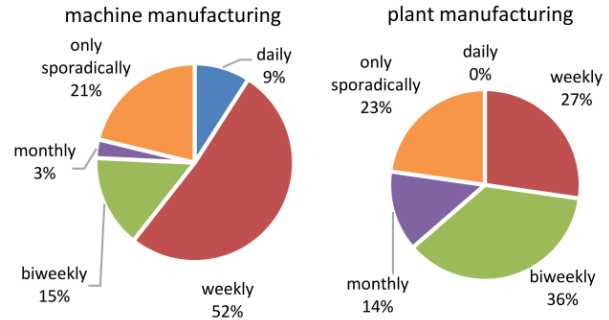

Fig. 11. Exchange frequency during the development phase. (#2.2)

### 6.2. Tools to support information exchange among different disciplines (RQ2.2)

Tools that support cross-disciplinary information exchange and their usage are analyzed in the following, focusing on cross-disciplinary variant and version management, but neglecting approaches for the more all-encompassing PLM.

*Usage of variant management tools (RQ2.2a)*

Unfortunately, still 44% of the companies don't use any tool to support variant management (#3.14), which is expressed in a red rating in Table IV (48% of all standardized machine manufacturers, 44% of special purpose machine manufacturers, 50% of PM companies and 37,5% of others). Derivation in between industries is at maximum 7% (chemical industry 41%, medical robotics 43%, mechanical engineering 48%). These results reflect the ease to build variants, which is of course much easier designing standardized machines than special purpose machines or plants. Regarding the used tools, the results are disillusioning, though. 32% use Excel (Macros), 11% a proprietary tool from a market leading automation company and 1% CodeSmith as configuration tool. 13% use other tools including different self-developed ones.

*Disciplines included in version management (RQ2.2b)*

Besides the management of variants, also versions of these variants over time need to be managed. Only 30% of the partic-ipating companies do not implement version management at all, but the tools companies use differ greatly (#3.10, Fig. 12).

TABLE IV
Evaluation of Research Question 2

| Research Questions | Detailed Research Questions | Findings | Evaluation results | Validity of results | Relevant Section |
|---|---|---|---|---|---|
| What are success factors for cross-disciplinary development? (RQ2) | How is cross-disciplinary development realized? (RQ2.1) | All relevant disciplines are taking part in meetings; frequency correlates with project duration | 🟢 | 🟢 | 6.1 |
| | Which tools are used for supporting information exchange among different disciplines? (RQ2.2) | SVN mostly for Software | 🔴 | 🟢 | 6.2 |
| | Usage of variant mgmt. tools (RQ2.2a) | 44% still do not use any tool for variant mgmt. | 🔴 | 🟢 | 6.2 1) |
| | Disciplines included in version mgmt. (RQ2.2b) | 70% of companies realize version mgmt. (SVN most used 29%), mostly for software (98%) and electrics / electronics (36%), other disciplines below 17% | 🔴 | 🟡 A | 6.2 2) |
| | Are cross-disciplinary modules used? (RQ2.3) | Only 21% use them by default | 🟡 | 🟡 SD | 6.3 |

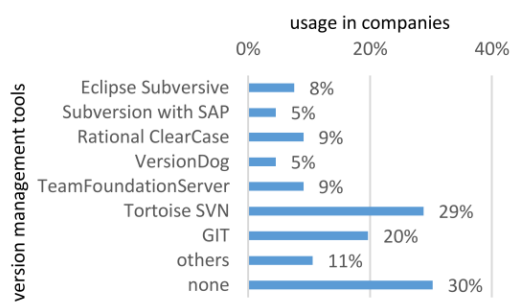

Fig. 12. Version management tools. (#3.10)

For purposefully updating the software of an operating plant, it is necessary to be aware of its exact setup regarding all involved disciplines. This exact setup is described by the variants and versions of all parts in use and the documentation of their combination. Precise updates, however, are contradicted by the software centric approach of version management, revealed by the analysis of the involved disciplines (#3.11, cp. Fig. 13). All companies realize version management for software, followed by electronics/electrical engineering. All other disciplines are below 17%, leading to a red rating of RQ2.2b (cp. Table IV). These low values reveal one possibility for improvement in an integrated engineering workflow for aPS.

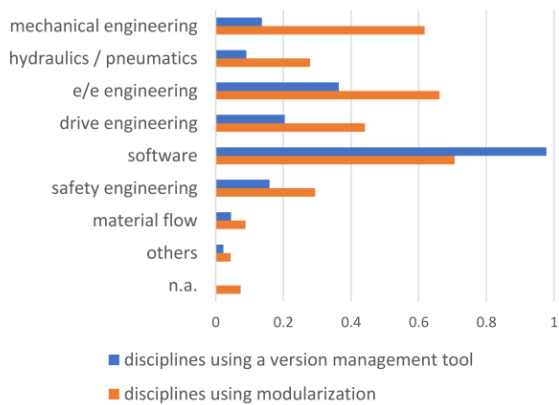

Fig. 13. Disciplines using a version management tool (#3.11) and disciplines using modularization. (#2.7)

Based on these insights, RQ2.2 is assessed as red in Table IV.

### 6.3. Cross-disciplinary modularization (RQ2.3)

One way to support the process of cross-disciplinary development is the usage of cross-disciplinary so-called mechatronic modules (#2.6). Until now only 21% of the companies use this approach by default, 49% partially, 30% do not use it at all. Fig. 13 identifies whether the goal of modularization is supported and implemented by all relevant disciplines (#2.7). Following the idea of mechatronics, mechanical engineering, electronics/electrical engineering and software are included with about 65%. Safety engineering, hydraulics and drives show lower percentages.

Summarizing RQ2.3, the usage of a cross-disciplinary modularization is with 21% quite low. This is indicated by an orange rating in Table IV.

## 7. Reusable Software Modules (RQ3)

Cross-disciplinary modularization as a means for system integration has been identified as a major driver for aPS [49, 50, 52, 59]. Since section 6 showed insufficient proficiency in industry, this section focuses on how such software modules can be realized in a reusable way. Therefore, it is not necessarily required to build mechatronic modules. In addition, discipline specific modules that are coupled in between the disciplines are one successful way (cp. Sec. 2.7). In the form of research questions, six detailed aspects address the underlying success factors for these reusable software modules.

Knowledge about compatibility between a software module and a specifically evolved machine or plant is crucial for successfully exchanging modules. Thus, influences among questions related to cross-disciplinary variant and version management are discussed first. Once a change in a software module is required, an appropriate release process helps managing and realizing the different variants and versions. One major prerequisite and especially important factor to be considered in this release process is the quality assurance of software modules before they are accepted in the module library. However, complexity of modules may disturb these measures and is therefore considered in form of additional information. To ease the engineering of reusable software modules and to cope with complexity, on the one hand, usage of implementation standards like hierarchical levels ISA 88 with appropriate standard functions and state machines (OMAC) for each module may be beneficial. On the other hand, code configuration from other engineering tools helps to decrease effort and time and increase quality, too.

### 7.1. Influence on and of variant and version management (RQ3.1)

The machines' degree of customization may be a disturbing factor for modularity, which is crucial for variant and version management of software. Therefore, the variables showing significant differences explained by the diverging use of machine specific control software are analyzed. A hypothesis test is conducted, comparing the results of companies with more than 50% machine specific control software and companies with less. Since normal distribution cannot be assumed, the Wilcoxon-Mann-Whitney-Test is performed. In case of binary variables a Chi-squared test or Fisher's exact test for small samples is used. The results, obtained with the software environment R, are significant at least at a level of 10% (cp. Table V, #3.17).

**TABLE V**
**Companies with more than 50% machine-specific control software code show dependencies with other variables**

| | |
|---|---|
| ↑ | usage of IEC 61131-3 AWL (#3.1) |
| ↑ | interfaces implemented as data exchange across global variables/flags (#3.6) |
| ↑ | Team Foundation Server as Version Management Tool (#3.10) |
| ↑ | source code hand-over to the customer (#5.3) |
| ↓ | n-axis-positioning rated as critical application (#1.14) |
| ↓ | degree of modularization (#2.7) * |
| ↓ | score standards for the implementation of software projects (#3.3) |
| ↓ | operating modes as standard function for modules (#3.7) |
| ↓ | amount of library blocks (#3.8) * |
| ↓ | score release process of library blocks (#3.9) |
| ↓ | disciplines using the version management tool (#3.11) |
| ↓ | usage of a variant management tool (#3.14) |
| ↓ | usage of automated configuration of the control software based on project templates (#3.15) |
| ↓ | usage of templates (#3.16) |

↑indicates increase, ↓indicates decrease
* indicates significance level of 5%, otherwise 10%

As expected, the results show the challenge of realizing modularization, standardization, variant and version management etc. for such a high degree of required customer specific adaptations. This means the higher the degree of customized functionality, the harder it is to achieve reusable modules.

To analyze the influencing factors on change tracking in more detail the companies using tracking of software changes in the block header are compared to companies with change tracking included in the version management tool. Further hypothesis tests with significance level of at least 10% reveal (cp. Table VI, #3.12) low utilization of modeling tools, cross-disciplinary mechatronic modules, interfaces, degree of modularization etc. for companies with block header. We conclude that mostly companies, which show poor values in other software maturity indicators, use this easy way of change tracking.

**TABLE VI**
**Change tracking (manual/history in the block header of software modules)**

| | |
|---|---|
| ↑ | interfaces implemented as data exchange across global variables/flags (#3.6) |
| ↓ | score utilization of modeling tools (#2.5) |
| ↓ | cross-disciplinary mechatronic modules (#2.6) |
| ↓ | degree of modularization (#2.7) |
| ↓ | score implementation of interfaces (#3.6) |
| ↓ | fault detection, positioning work (NC) and operating modes as standard function for modules (#3.7) |
| ↓ | amount of library blocks (#3.8) |
| ↓ | usage of a version management tool (#3.10) |
| ↓ | usage of automated configuration of the control software based on project templates (#3.15) |
| ↓ | score measures for quality assurance (#4.1) |

As the most appropriate and mature variant management is assumed to ease software management, the dependencies in between the variant management tool (#3.14) and any other variable of the questionnaire are analyzed for significant dependencies (cp. Table VII). Further hypothesis tests with significance level of at least 10% were conducted. None of the tools shows a significant positive impact on all factors meaning that the usage of this tool would support high values in these factors. For example, Excel is related to a higher frequency in exchange during engineering and better processes and continuous integration. Both the proprietary tool from an automation company (pAT) and Excel are related to a higher usage of implementation standards like ISA 88 or 95 (+ in green color #3.3 in Table VII). Again, it is disillusioning and shows the potential for improvement to realize that companies using Excel as variant management tool, reach overall the best values (average of maximum scores in entire questionnaire: Excel: 57%, pAT: 51%, None: 47%). A negative impact shows that in case Excel is used, the number of disciplines included shows lower values (average of maximum scores for question #2.1, all companies: 70%; companies using Excel: 56%), as well as the maturity of the version management tool (average of maximum scores for question #3.10, all companies: 66%; companies using Excel: 50%). The orange rating of RQ3.1 in Table VIII mirrors these insights.

**TABLE VII**
**Dependencies between selected questions and different variant-management tools (green / + -positive impact, red / - -negative impact)**

| Evaluation criterion | None | Excel | pAT |
|---|---|---|---|
| Companies in Plant Manufacturing (#1.3) | + | - | + |
| Number of disciplines exchanging (#2.1) | + | - | + |
| Frequency of exchange (#2.2) | - | + | - |
| Cross-disciplinary mechatronic modules (#2.6) | - | + | |
| Implementation standards (#3.3) | - | + | + |
| Software standard functions (#3.4) | - | - | + |
| Creation and release of a library block (#3.9) | | + | - |
| Version management tool (#3.10) | + | - | - |
| Continuous Integration (#3.13) | - | + | |
| Templates (#3.16) | - | + | |

## 7.2. Release process of library blocks (RQ3.2)

The release process of library blocks has a great impact on the software quality of libraries. This is especially relevant in PM as the first test is often not conducted until the plant is built. In case of new machine or plant generations, both software and system are further developed on different sites in parallel. This results in different variants and versions that need to be evaluated and merged before being accepted as standard modules for the library. Only then can they be used as a basis for the next machines or plants. A well-designed release process is marked by different employees developing and testing/releasing a library block and ideally a parallel interception. Such a well-designed release process is depicted in Fig. 14.

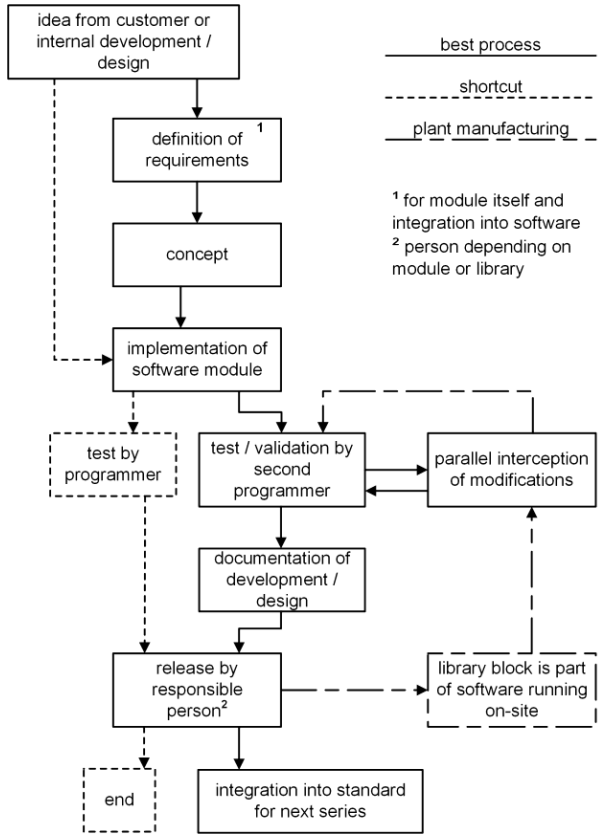

Fig. 14. Release process for library blocks. (#3.9)

Among the companies participating in the survey, a majority uses a defined release process for library blocks. However,

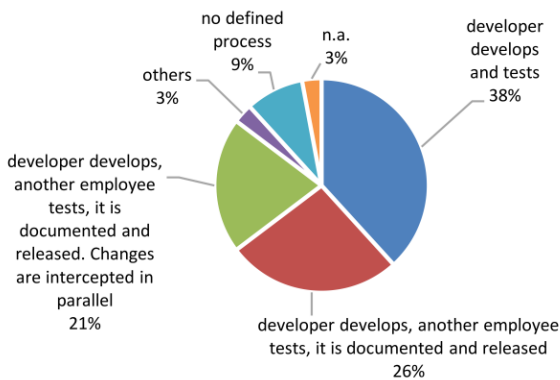

Fig. 15. Release process for library blocks. (#3.9)

in 38% of these companies, the library block's developer is also the one testing the module. While 26% of companies rely on two employees for development and testing/release, only 21% implement the ideal release process. This distribution is depicted in Fig. 15. Considering the possible improvements concerning the release process, RQ3.2 is rated orange in Table VIII.

## 7.3. Measures for quality assurance (RQ3.3)

A variety of measures exists for quality assurance. These range from code reviews to testing at the desk or the machine to automated testing. A valuable insight is, that the earlier faults are discovered the less costly their correction. An overview of the measures implemented by the different companies and their prevalence is depicted in Fig. 16. It is astonishing, that 42% of companies apply automated software testing. However, 10% of the companies still rely solely on testing at commissioning (thus orange rating in Table VIII).

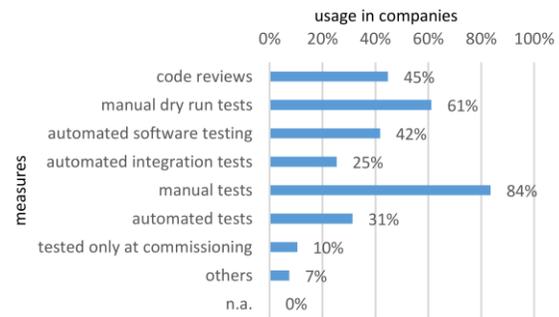

Fig. 16. Measures for quality assurance. (#4.1)

Apart from the measures used for quality assurance, it was also analyzed which scenarios are tested, cp. Fig. 17. Requirement coverage is tested by the majority of companies, i.e. every specified scenario is tested. Almost as many conduct acceptance tests for good behavior only. Still a third of all companies test towards FDA, which is typical for machines and plants intended for the food industry. Code coverage, however, is realized by only 15%, which may be due to its complexity and the necessary effort for testing all possible scenarios.

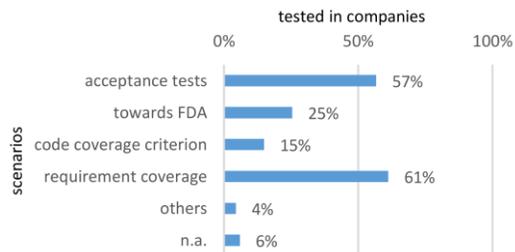

Fig. 17. Tested scenarios. (#4.2)

## 7.4. Complexity as an important characteristic (RQ3.4)

The surveyed companies unanimously regard complexity as a highly relevant characteristic of control software projects (#1.12, #1.13), as complexity correlates to the necessary effort of realizing a desired function. Knowledge about the complexity of software enables e.g. properly assigning resources and implementing appropriate quality controls. Adequately measuring complexity turned out to be a challenge, though. Typical indicators such as the number of CPUs, POUs or lines of code fall short of properly illustrating complexity. A more construc-tive but less objective way of assessing complexity would be to identify the most critical task, or conduct code analyses as done by Vogel-Heuser et al. [7], which requires far more effort. Concluding, there exist no simple but effective measure for complexity. This impedes focusing on crucial elements and leads to a red rating of RQ3.4 in Table VIII.

TABLE VIII
Evaluation of Research Question 3

| Research Questions | Detailed Research Questions | Findings | Evaluation results | Validity of results | Relevant Section |
|---|---|---|---|---|---|
| What are success factors for reusable software modules? (RQ3) | How big is the variant and version management tools' influence? (RQ3.1) | Potential for improvement: companies using Excel as variant management tool reach overall the best values | 🟡 | 🟡 A | 7.1 |
| | How are library blocks released? (RQ3.2) | Defined release process for library blocks (88%), only 21% implement the ideal release process | 🟡 | 🟢 | 7.2 |
| | How is quality assured? (RQ3.3) | 10% of the companies still rely solely on testing at commissioning | 🟡 | 🟢 | 7.3 |
| | What is the influence of complexity? (RQ3.4) | Adequately measuring complexity is still a challenge | 🔴 | 🟢 | 7.4 |
| | Which standard functions are most used? (RQ3.5) | Common: Fault detection (65%), diagnosis (50%) and modes of operation (46%) | 🟡 | 🟡 SD | 7.5 |
| | Is code configuration from engineering tools realized in industry? (RQ3.6) | 63% of MM and 45% of PM don't apply automatic configuration | 🔴 | 🟢 | 7.6 |

### 7.5. Standard functions (RQ3.5)

Standard functions embedded in every software module support software architectures and ease changes and adaptation. Therefore, the application of standard functions in software modules is analyzed (#3.7, cp. Fig. 18). Fault detection and diagnosis as well as modes of operation are often implemented (orange rating in Table VIII). Reasons and obstacles should be elaborated in future work.

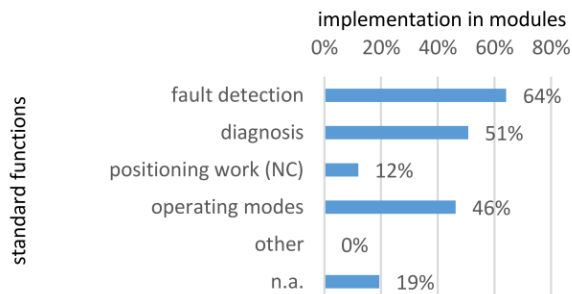

Fig.18. Module standard functions. (#3.7)

### 7.6. Code configuration from engineering tools (RQ3.6)

Software templates may be used to efficiently develop the automatic configuration of an application. As input, information from other engineering disciplines' tools like proprietary electrical engineering tools (pECAE), UML models or even Excel is used (cp. Table IX, #3.15). The results are shocking, as 63% of MM and 45% of PM do not apply automatic configuration at all (red rating in Table VIII). Even though it should be much easier to apply such configuration in MM, they use it less often than in PM. As input information, Excel and a market leading pECAE tool both have a share larger than 35% in PM. In MM, a UML based approach embedded in IEC 61131-3 already gains 23%, which is promising for future improvements for MDE based approaches. In MM, high-level programming languages are used more often than in PM, as we know from different interviews. Therefore, proven MDE tool chains using UML and code generation ease the application of the MDE approach requiring further investigations with a larger number of companies.

Table IX
Evaluation of tools for code generation and automated configuration of control software based on project templates (#3.15)

| | | PM | MM |
|---|---|---|---|
| Tools for Code Creation | pECAE | 37.5 % | 8 % |
| | Excel (Macros) | 37.5 % | 38 % |
| | UML Plug In | 6 % | 23 % |
| Usage of Automated Configuration | Standard method | 14 % | 5 % |
| | Partially | 41 % | 32 % |
| | No | 45 % | 63 % |

## 8. Major Influencing Factors and Inferred Necessary Measures

An overview of the industry-specific maturities in the different influencing factors is presented in Fig. 19. On the one hand, the depicted influencing factors are selected to compare the results of the new survey SWMAT4aPS+ with its 68 companies and the prior SWMAT4aPS results from 16 companies [7]. The similarity of the results proves the validity of the approach. On the other hand, further relevant questions are included, e.g. on the success factors for reusable software modules in the different application domains represented by RQ3.6. Technologies applied in medical robotics are the most advanced, but chemistry and mechanical engineering are not far behind. The overall maturities for the disciplines chemistry, mechanical engineering, and medical robotics are 2.81, 2.64, and 2.85 respectively compared to the maximum score of 5, whereas multiple selections concerning the classification were allowed in the questionnaire. The average maturity of all companies that participated in the survey was 2.68. This includes also companies that could not be assigned to one of the three categories specified above. Concluding, the factors with the biggest influence on the overall maturity of companies are identified. For this purpose, the correlation between the companies' scores for individual questions with the overall rating is analyzed.

Major influencing factors were identified, namely the measures used for quality assurance (#4.1, RQ3.3), the procedure for releasing library blocks (#3.9, RQ3.3), interdiscipli-nary usage of a version management tool (#3.11, RQ2.2), the degree to which simulation is used for testing (#4.3, RQ1.2) and the use of an automated configuration of the control software (#3.15, RQ3.6). These factors, their correlations to the overall maturity and the corresponding average scores are listed in Table X.

Table X
Major Influencing Factors

| question (#) | correlation to overall maturity | avg. score | avg. score [7] |
|---|---|---|---|
| measures for quality assurance (#4.1) | 0.695 | 3.01 | 4.06 |
| library block release process (#3.9) | 0.642 | 2.29 | 2.50 |
| version management tools (#3.11) | 0.611 | 2.18 | 3.33 |
| testing by use of simulation (#4.3) | 0.592 | 2.22 | 2.60 |
| automated configuration of control software (#3.15) | 0.571 | 1.38 | 1.46 |

As Table X shows, the average values for these important questions were all significantly small, i.e. below 2.5, with the exception of #4.1. These uncovered deficits may be used as levers to greatly improve a company's individual maturity. When comparing these results with the ones from previous work [7], it becomes apparent, that the previous values are all higher, but the tendency is the same. The higher values within [7] may be explained by the fact that the companies in the first case study were selected because they already cooperated with academia. In addition, a number of 16 companies is not representative. Explanations of the influencing factors' importance and suggestions for improvement are presented in the following.

A high rating in quality assurance (#4.1, RQ3.3) indicates that the company has identified quality as a major driver for success. Such a mindset greatly influences the company's overall maturity. This seems to be widely recognized in industry, as it is also the critical factor best fulfilled. However, unused potentials lie within automated testing, which is usually still not sufficiently realized.

While release processes for library blocks (#3.9, RQ3.3) are almost a standard in industry, in most cases they are not well designed (cp. Sec. 7.2, RQ3.2). This is problematic as such release processes facilitate managing and realizing different variants and versions. In addition, quality assurance is part of the release process. Hence, a well-defined release process influences the degree to which quality assurance is realized, too. The combination of these factors explains the high influence of the release process on the overall maturity.

Version management is usually realized for software development. This is not the case for the other disciplines involved in the development, though. Considering the importance of cross-disciplinary cooperation in the field of MM and PM, the implementation of such a cross-disciplinary version management is highly desirable (#3.11, RQ2.2). This reflects in its importance for the overall maturity.

The degree to which simulation is used for testing (#4.3, RQ1.2) also has an evident influence on the overall maturity. This is because it enables testing and quality checks already in early phases before start-up. Simulation can be realized by use of modeling tools or simulation languages i.e. Matlab/Simulink.

Finally, automated configuration of control software (#3.15, RQ3.6) should be supported through project templates and appropriate tools as delineated in section 7.6. Automated configuration is this important for the overall maturity because it decreases effort and time of the development and increases the software's and thus the whole system's quality.

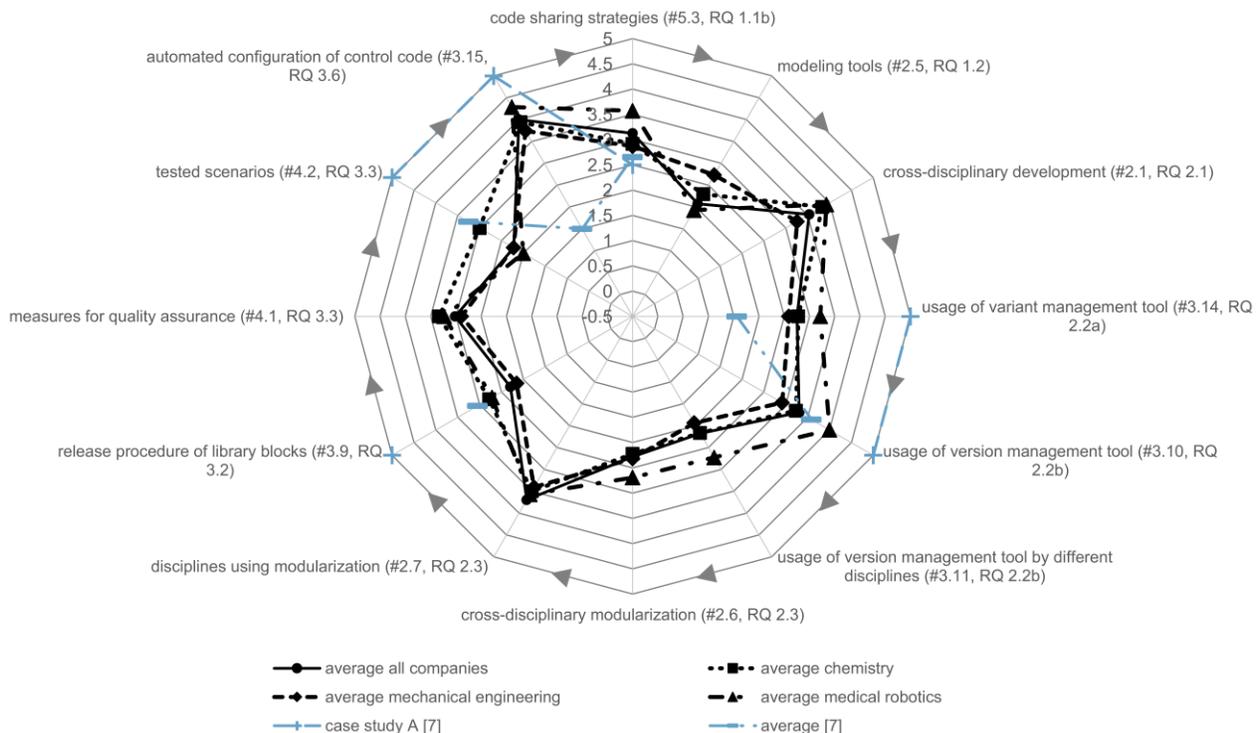

Fig. 19. Industry-specific maturities for different influencing factors.

## 9. Conclusion and Outlook

Using SWMAT4aPS+ with its self-assessment questionnaire, we gained deeper insights in the state of the art in software engineering of aPS as an enabler to evolve aPS permanently. Permanent evolution of software is a prerequisite to allow flexibility for new products in Industry 4.0 also after handover to the customer, the operating company. SWMAT4aPS+ addressed three main research questions (RQ) based on a questionnaire answered by 68 German companies in the field of aPS delivering to the world market:

- enabling exchange of aPS software modules after acceptance test (RQ1)
- success factors for cross-disciplinary development (RQ2)
- success factors for reusable software modules (RQ3)

Several influencing factors were identified for these three categories. The maturity in the different sub research questions (RQ1.1 to RQ3.6) is still low in comparison to other domains like automotive (cp. Table III, IV, and VIII, orange and red dots in column "Evaluation results"). All the results could be validated apart from the ones marked orange or red in column "Validity of results".

It is evident that companies do not have sufficient knowledge about variants and versions of software for updating due to uncertainty of the software status implemented at the customer's site (RQ1.1). Another weakness was identified regarding the tool support for information exchange among different disciplines (RQ2.2) and variant and version management. Regarding the identification of success factors for reusability of software modules all considered factors show room for further improvement (RQ3.1 to RQ3.6). This mirrors the necessity for further research on the obstacles for companies to improve their processes on the one hand and on the improvement of approaches and tools adequate for software engineering in the aPS domain on the other hand.

The presented results provide 19 criteria (detailed RQs) for companies to benchmark their engineering, maintenance and service departments regarding maintainable and evolvable software modules in comparison to other companies in the same field. To implement the lessons learned from the first questionnaire [7], the option to provide free text, which was then used to check for inconsistencies within the answers, was included (cp. Sec. 3.1). No inconsistencies could be identified, but the free text leads to additional insights concerning tools used in industry. Nevertheless, there may still exist hidden relations not yet revealed, due to weaknesses in the questionnaire and/or limited number and sample of companies included in the questionnaire. An issue already identified within the first application of the questionnaire [7] is the investigation of country-related specialties. Even though a broad variety of different companies was included this time, an analysis according to countries was not possible. This shall be realized in future work by providing the questionnaire in English via a web based access. So far, the only international data available stems from an Austrian and an Italian company. It shows similar results for the survey, which can be understood as an indicator for the survey's external validity. This, however, will need to be confirmed in future work, too.

During the analysis of the questionnaire's results, some lessons were learned for a further development of SWMAT4aPS+:

- Concerning the usage of high-level programming languages and more sophisticated control platforms, the recent questionnaire does not distinguish between control, SCADA, HMI, MES and other IT related applications, which should be included in future surveys (cp. Sec. 5.3).
- Further distinction of MM companies into standardized machine and special purpose machine manufacturers, which should have improved in comparison to the previous results [7], did not work due to multiple answers in one question. This was not expected, but can be explained by the mix of machine types delivered by most companies (cp. Sec. 4).
- Unfortunately, the relations of used controllers, industrial sector and programming paradigm could not be revealed out of the data and need to be further examined in future research, too (cp. Sec. 5.3).
- The low values for disciplines apart from SW using a version management tool reveal a huge potential for improvement in an integrated engineering workflow for CPPS but also the requirement for more appropriate tool support (cp. Sec. 6.2).
- Reasons for and obstacles to using standard functions should be researched in detail in future work (cp. Sec. 7.5).
- Automated software testing showed (with 42%) unexpected high values compared to [7] and experiences with different companies in six research projects on testing in automation. It is assumed that the answers refer mostly to unit tests of modules, but the question needs to be refined (cp. Sec. 7.3).

A revised questionnaire in German and English is already available online. It will deliver additional insights into some of the remaining open questions as well as into electrical engineering aspects. In parallel, efforts should be made to develop metrics for appropriately assessing complexity of software library modules in the aPS domain. As a subsequent step, the authors also aim at comparing and combining the results of the already conducted surveys. This allows, among others, to further investigate the influence of modularity on maintainability and evolvability.

It is assumed that the approach will also be applicable for embedded software in construction machines, agricultural machines and other Embedded Systems using PLC-based control nodes or IEC 61131-3 programming environments. Such Embedded Systems also require real-time behavior and reliability. For these domains, the requirements and constraints are similar aside from the number of produced products and their individual adaptations.

## Acknowledgment

We thank all companies who answered the questionnaire. This study was partially funded by DFG (Deutsche Forschungsgemeinschaft) through the fund SPP 1593 project DOMAIN and MODEMICAS.